\documentclass[9pt,twocolumn,twoside]{osajnl}

\journal{optica} 

\setboolean{shortarticle}{false}

\usepackage{graphicx,float}
\usepackage{braket}
\usepackage{siunitx}
\usepackage{appendix}
\usepackage{amssymb}
\usepackage{mathrsfs}
\usepackage{appendix}
\usepackage{slashed}
\usepackage{physics}
\usepackage{mathtools}
\usepackage{amsmath}
\usepackage{bbold}
\usepackage{xcolor}
\usepackage[mathscr]{euscript}
\usepackage{babel}
\usepackage{tabularx}
\usepackage{changes}

\newcommand{\beq}{\begin{equation}}
\newcommand{\eeq}{\end{equation}}
\renewcommand{\d}{{\rm d}}

\title{Temporal trapping: a route to strong coupling and deterministic optical quantum computation}

\author[1,*]{Ryotatsu~Yanagimoto}

\author[1,2]{Edwin~Ng}

\author[1,2]{Marc~Jankowski}

\author[1]{Hideo~Mabuchi}

\author[2,3,$\dagger$]{Ryan~Hamerly}

\affil[1]{E.\,L.\,Ginzton Laboratory, Stanford University, Stanford, California 94305, USA}
\affil[2]{Physics \& Informatics Laboratories, NTT Research, Inc., Sunnyvale, California 94085, USA}
\affil[3]{Research Laboratory of Electronics, MIT, 50 Vassar Street, Cambridge, MA 02139, USA}
\affil[*]{Corresponding author: ryotatsu@stanford.edu}
\affil[$\dagger$]{Corresponding author: rhamerly@mit.edu}

\dates{Received 18 August 2022; revised 13 October 2022; accepted 14 October 2022; published 17 November 2022}

\doi{\url{https://doi.org/10.1364/OPTICA.473276}}

\begin{abstract}
The realization of deterministic photon-photon gates is a central goal in optical quantum computation and engineering. A longstanding challenge is that optical nonlinearities in scalable, room-temperature material platforms are too weak to achieve the required strong coupling, due to the critical loss-confinement tradeoff in existing photonic structures. In this work, we introduce a novel confinement method, dispersion-engineered temporal trapping, to circumvent the tradeoff, paving a route to all-optical strong coupling. Temporal confinement is imposed by an auxiliary trap pulse via cross-phase modulation, which, combined with the spatial confinement of a waveguide, creates a “flying cavity” that enhances the nonlinear interaction strength by at least an order of magnitude. Numerical simulations confirm that temporal trapping confines the multimode nonlinear dynamics to a single-mode subspace, enabling high-fidelity deterministic quantum gate operations. With realistic dispersion engineering and loss figures, we show that temporally trapped ultrashort pulses could achieve strong coupling on near-term nonlinear nanophotonic platforms. Our results highlight the potential of ultrafast nonlinear optics to become the first scalable, high-bandwidth, and room-temperature platform that achieves a strong coupling, opening a new path to quantum computing, simulation, and light sources.
\end{abstract}

\setboolean{displaycopyright}{true}

\begin{document}

\maketitle
\section{Introduction}
Photons are ideal carriers of quantum information, enjoying minimal decoherence even at room temperature, and propagating long distances with low loss at high data rates. These advantages render optics essential to quantum key distribution~\cite{Yin2016}, networking~\cite{Kimble2008}, and metrology \cite{LIGO2013,Treps2002}, and have led to significant progress towards optical quantum computation~\cite{Obrien2009,Asavarant2019,Obrien2007}. The main challenge to the latter lies in realizing on-demand entangling gates between optical qubits, in light of the weak photon-photon coupling in most materials.  The dominant paradigm---linear optical quantum computing (LOQC)---circumvents this problem via the inherent nonlinearity of measurements~\cite{Knill2001}, but as the resulting gates are probabilistic~\cite{Knill2002}, LOQC relies on the creation of entangled ancillae~\cite{Knill2001} or cluster states \cite{Raussendorf2001, Nielsen2004,Reimer2019}, which suffer from large resource overheads in terms of the number of photons and detectors per gate \cite{Slussarenko2019, Li2015, Kok2007, Rudolph2017}.

\begin{figure*}[bt]
    \includegraphics[width=0.98\textwidth]{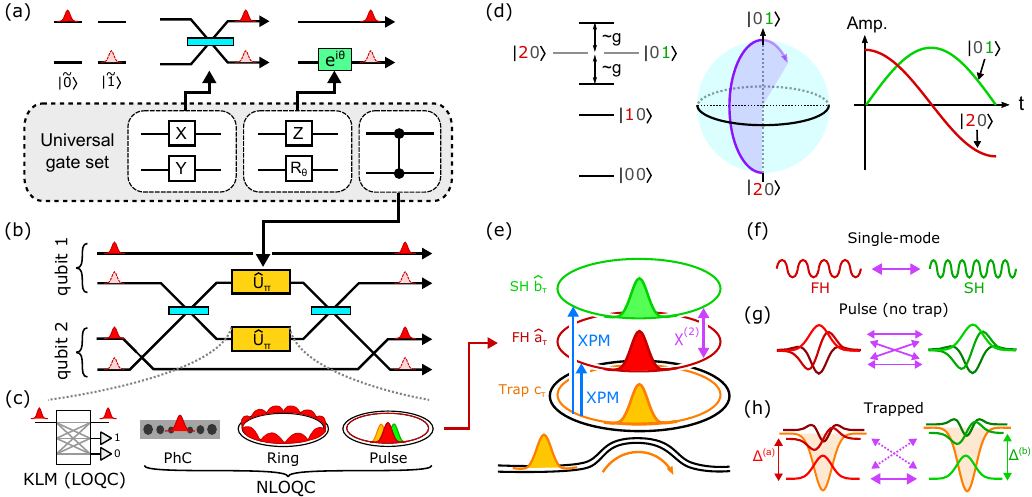}
    \caption{Universal QC is realized on a dual-rail qubit basis with (a) single-qubit gates based on passive linear optics, and (b) a CZ gate constructed from a Kerr-phase interaction $\hat{U}_\pi$ inside a Mach-Zehnder interferometer. (c) Potential realizations of $\hat{U}_\pi$ in LOQC and NLOQC. (d) $\chi^{(2)}$-mediated $\hat{U}_\pi$ gate: coupling between the FH state $\ket{2\,0}$ and the SH state $\ket{0\,1}$ leads to Rabi oscillations, imparting a nonlinear phase shift on the signal field. (e) Temporal trap: the $\chi^{(2)}$ interaction between FH and SH fields is enhanced when confined to ultrashort pulses through trap-pulse XPM.  Untrapped dynamics are either (f) CW and single-mode or (g) pulsed and multimode, depending on the dispersion.  (h) Temporal trapping imposes single-mode dynamics by breaking the degeneracy between trapped and untrapped modes, the former protected by an energy gap $\Delta$.}
    \label{fig:f1}
\end{figure*}

The inherent difficulty of probabilistic gates has fueled sustained interest in so-called nonlinear-optical quantum computing (NLOQC), where deterministic gate operations are implemented coherently through a nonlinear-optical interaction \cite{Chuang1995,Langford2011}. Here, high-fidelity gates are possible in the strong-coupling regime when the nonlinear interaction rate $g$ exceeds the decoherence rate $\kappa$, i.e., $g/\kappa\gg 1$. Strong coupling is readily achieved in cavity QED, where resonant two-level systems such as atoms mediate strong optical nonlinearities~\cite{Yoshie2004, Birnbaum2005, Englund2007,Hacker2016,Boca2004}, but such systems require vacuum and/or cryogenic temperatures, and challenges with fabrication, yield, and noise remain daunting despite decades of research. By contrast, bulk material nonlinearities such as $\chi^{(3)}$ and $\chi^{(2)}$ are robust, scalable, and room-temperature, but the optical interaction is much weaker, imposing very demanding requirements on the optical loss (quality factor $Q$) and confinement (mode volume $V$). Moreover, to support nonlinear interactions among multiple frequency bands, e.g., in $\chi^{(2)}$ systems, one has to overcome the challenge of realizing high-$Q$ resonances separated a large frequency, for which guided-wave (e.g.\ ring, disk) resonators are favorable options compared to photonic crystal cavities. Great progress has been achieved to this end in ultra-low-loss thin-film LiNbO$_3$ (TFLN)~\cite{Zhang2017,Lu2020} and indium gallium phosphide (InGaP) nanophotonics~\cite{Zhao2022}, which has rendered plausible a near-strong coupling regime $g/\kappa\sim1$ with ring resonators in the near future. Even with these developments, however, $g/\kappa\gg 1$ remains a challenge owing to the ring's large mode volume, as the axial dimension remains unconfined. To reach strong coupling, field confinement in the transverse dimensions is not enough. We also need a means to confine light in the third direction---time.

This paper introduces the {\it temporal trap}, a nonlinear-optical mechanism to confine light in time as well as space. To facilitate trapping, a strong non-resonant ``trap pulse'', which co-propagates with the target fields, introduces a nonlinear phase shift through cross-phase modulation (XPM). Analogous to an optical soliton \cite{AgrawalBook}, the trap pulse creates a flying photonic cavity that supports a bound mode formed by the competition between dispersion and nonlinearity, with a mode volume reduced by the trap duty cycle. With appropriate dispersion engineering~\cite{Jankowski2021-review}, the bound mode is strongly detuned from the remaining cavity degrees of freedom, ensuring single-mode dynamics that circumvent the inherent challenges of pulsed nonlinear quantum gates highlighted in Ref.~\cite{Shapiro2006, Shapiro2007}. As a result, we show that high-fidelity two-qubit entangling gate (i.e., controlled-Z gate) operation is possible, providing a roadmap to fully deterministic NLOQC. The tight temporal confinement also significantly increases nonlinear coupling strength, with $g/\kappa\gtrsim 10$ plausible for realistic nonlinearities and propagation losses on TFLN photonics. While we focus on $\chi^{(2)}$ systems as a case study in this work, our proposal is generic and compatible with existing proposals in NLOQC using $\chi^{(3)}$ nonlinear interactions as well~\cite{Milburn1989,Chuang1995,Langford2011}, where it both provides a means to resolve the otherwise unavoidable multimode interactions and also enhances nonlinear coupling strength. Additionally, our prescription using temporal traps supports time multiplexing~\cite{Inagaki2016,Takeda2017}, enabling significant parallelism in a single cavity.

\section{Optical Quantum Computing in a Temporal Trap}

Single-photon qubits are a leading approach for optical quantum computation~\cite{Obrien2007}.  The dual-rail basis, which encodes a state in polarization~\cite{Crespi2011}, time-bin~\cite{Humphreys2013}, or path~\cite{Qiang2018,Obrien2003}, is a particularly attractive choice, since all single-qubit gates reduce to linear optics (Fig.~\ref{fig:f1}(a)).  To complete the gate set, we also need a two-qubit entangling gate, e.g., a controlled-Z (CZ) gate.  The most common prescription, shown in Fig.~\ref{fig:f1}(b), implements CZ with a Mach-Zehnder interferometer (MZI) that encloses a Kerr-phase interaction:
\begin{align}
    \label{eq:mapping}
    \hat{U}_{\pi}\bigl[c_0\ket{0}+c_1\ket{1}+c_2\ket{2}\bigr]=c_0\ket{0}+c_1\ket{1}-c_2\ket{2},
\end{align}
where $\ket{n}$ represents the $n$-photon Fock state. This circuit exploits the Hong-Ou-Mandel effect~\cite{Hong1987} to ensure that two photons are incident on the $\hat{U}_\pi$ gate only when the qubits are in the logical state $\ket{\tilde{1}\,\tilde{1}}$, implementing the $\pi$-phase shift exclusively for this state.

To implement $\hat{U}_\pi$ (Fig.~\ref{fig:f1}(c)), one can employ the Knill-Laflamme-Milburn (KLM) scheme, which forms the basis for LOQC~\cite{Knill2001, Knill2002}.  KLM suffers from a low success probability of $2/27$ for the CZ gate, and deterministic operations require the preparation of an initial highly entangled state, e.g., a cluster state~\cite{Raussendorf2001, Nielsen2004}, at significant overhead~\cite{Li2015}. In light of these difficulties, here we focus on NLOQC, which aims at deterministic gate operations using coherent nonlinear dynamics~\cite{Chuang1995,Langford2011}. For instance, unitary evolution under a single-mode Kerr nonlinearity $\hat{H}_{\text{gate}}=\frac12\chi\hat{a}^{\dagger 2}\hat{a}^2$ for time $t_\pi=\pi\chi^{-1}$ implements $\hat{U}_\pi$. In this work, we instead consider a single-mode degenerate $\chi^{(2)}$ Hamiltonian
\begin{align}
\label{eq:single-mode-chi2}
    \hat{H}_{\text{gate}}=\frac{g}{2}(\hat{a}^2\hat{b}^\dagger+\hat{a}^{\dagger2}\hat{b}),
\end{align}
where $\hat{a}$ and $\hat{b}$ are annihilation operators for the fundamental (FH) and second harmonic (SH) modes, respectively. As shown in Fig.~\ref{fig:f1}(d), the Hamiltonian \eqref{eq:single-mode-chi2} mediates interactions between the two-photon FH state $\ket{2\,0}$ and the single-photon SH state $\ket{0\,1}$ with coupling strength $g>0$, resulting in a Rabi oscillation between these two states. Importantly, for an initial state of $\ket{2\,0}$, the system oscillates back to the same state after a period of $t_\pi=\sqrt{2}\pi g^{-1}$ with \emph{an opposite sign}, i.e., $-\ket{2\,0}$. As a result, for an initial FH state of $c_0\ket{0}+c_1\ket{1}+c_2\ket{2}$ and a vacuum pump state, unitary evolution under \eqref{eq:single-mode-chi2} for time $t_\pi$ implements $\hat{U}_\pi$ deterministically. Such a nonlinear-optical implementation of $\hat{U}_\pi$ is also considered in Refs.~\cite{Langford2011,VanDevender2007,Irvine2006,Majumdar2013}, which motivates us to employ this as a reference protocol for evaluating the performance of our proposal.

Now, the problem of implementing a CZ gate reduces to the realization of the single-mode $\chi^{(2)}$ Hamiltonian \eqref{eq:single-mode-chi2} with strong coupling, for which we sketch three possible realizations in Fig.~\ref{fig:f1}(c): a photonic-crystal cavity (PhC), a micro-ring resonator, and our proposed scheme using an ultrashort pulse. For resonators, the cooperativity figure of merit $g/\kappa = g/\sqrt{\kappa_a \kappa_{b}}$ depends on the $Q$ factor and mode volume as follows:
\beq
	\frac{g}{\kappa} = \sqrt{\frac{4\pi \hbar c d_{\rm eff}^2}{n^3\epsilon_0 \lambda^4} \frac{Q_a Q_{b}}{\tilde{V}}}, \label{eq:g_kappa}
\eeq
where $n$ is the refractive index of the medium, $\tilde{V} = V/(\lambda/n)^3$ is the normalized volume, with $V = \bigl|n^3 \int{E_{b}^* (E_a)^2 \d^3\vec{x}}\bigr|^{-2}$ defined in terms of the mode overlap integral between FH and SH modes. Effective quadratic susceptibility of the medium $d_\text{eff}$ is related to the native quadratic susceptibility $d_{33}$ via $d_\text{eff}=d_{33}$ and  $d_\text{eff}=(2/\pi)d_{33}$ for critical phase matching and quasi phase matching, respectively (See Supplement 1 for details).

\newcolumntype{L}[1]{>{\hsize=#1\hsize\raggedright\arraybackslash}X}%
\newcolumntype{R}[1]{>{\hsize=#1\hsize\raggedleft\arraybackslash}X}%
\newcolumntype{C}[1]{>{\hsize=#1\hsize\centering\arraybackslash}X}%

\begin{table}[tb]
\begin{center}
\begin{tabularx}{1.0\columnwidth}{C{1.10}|C{0.75}C{0.75}|C{0.85}|C{1.4}|C{0.9}|C{1.1}}
\hline\hline
& $Q_{a}$& $Q_{b}$ &
$\tilde{V}$ & $d_{\rm eff}$ &
$g/\kappa$ & Modes \\ \hline
PhC$^{*}$ & $10^6$ & $10^3$ & 1 & 33 pm/V & 0.03 & 1 \\
Ring$^\dagger$ & $10^7$ & $10^7$ & $2000$ & 21 pm/V & 0.1 & 1 \\
Pulse$^\ddag$ & $10^7$ & $10^7$ & $40$ & 21 pm/V & $0.8$ & $\gg 1$ \\  \hline\hline
\end{tabularx}
\end{center}
\caption{Typical estimates of $Q$, $V$, and cooperativity for competing confinement mechanisms.  LiNbO$_3$, $\lambda = 1.55$~$\mu$m.  $^*$Doubly-resonant PhC based on intersecting nanobeams, BIC, or nanopillars \cite{Rivoire2011, Minkov2019, Lin2016, Lin2017}.  $^\dagger$Ring circumference 2~mm, quasi phase-matched $d_{\rm eff} = (2/\pi)d_{33}$, loss $\alpha = 3$~dB/m, and $Q=5\times10^6$ at $\lambda=\SI{1.59}{\micro m}$ \cite{Zhang2017}.  $^\ddag$Pulse of width $\SI{100}{fs}$, dispersion engineered waveguide.}
\label{tab:t1}
\end{table}

Table~\ref{tab:t1} reveals the tradeoff between $Q$ and $V$ in resonator design. In terms of their generic properties, a PhC cavity leverages a wavelength-scale mode volume $V \lesssim (\lambda/n)^3$ with modest $Q\sim 10^6$ ($Q \sim 10^7$ is in principle possible, but at low yield \cite{Quan2011, Asano2019, Sekoguchi2014, Minkov2014, Asano2017, Dodane2018, Taguchi2011}).  However, as PhCs rely on Bragg scattering for confinement, simultaneous resonance of octave-spanning modes is very difficult, leading to lower quality factors $Q \lesssim 10^4$ at the SH~\cite{Rivoire2011, Minkov2019, Lin2016, Lin2017}.  On the other hand, the light in ring resonators is guided by total internal reflection, a geometric effect that is only weakly wavelength-dependent.  Therefore, rings can readily resonate modes spanning an octave, with $Q$ factors limited only by waveguide loss.  With ion-sliced TFLN, losses of 3~dB/m ($Q = 10^7$) have been achieved \cite{Zhang2017}, and there is a pathway to reach $Q = 10^8$ with process improvements \cite{Shams2021, Gao2021, Gao2022}, which is close to the bulk material limit \cite{Ilchenko2004, Serkland1994, Savchenkov2004, Schwesyg2010}. For the Kerr effect, PhC cavities offer better performance; however, the native nonlinearity is still too weak in standard materials to observe strong coupling with reasonable cavity designs (see Supplement 1). More sophisticated engineering methods, e.g., coherent photon conversion~\cite{Langford2011,Ramelow2019}, could provide further enhancement to the nonlinearities on $\chi^{(3)}$ platforms. For $\chi^{(2)}$, ring resonators are the superior option. Recent experiments have demonstrated $g/\kappa \sim 0.01$ on ultra-low-loss TFLN ~\cite{Lu2020} and InGaP~\cite{Zhao2022} micro-ring resonators; however, the strong-coupling regime $g/\kappa\gg 1$ remains challenging due to the ring's large mode volume.

\begin{figure*}[tb]
\begin{center}
\includegraphics[width=1.00\textwidth]{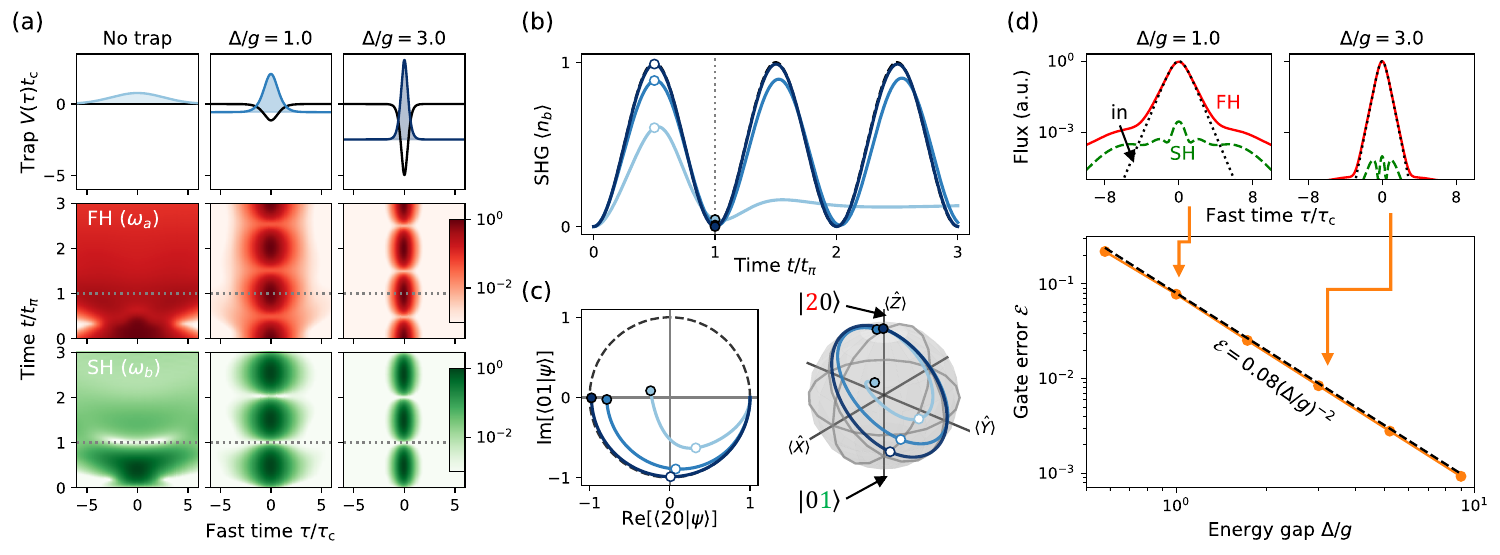}
\caption{Two-photon Kerr-phase gate $\hat{U}_\pi$ with and without temporal trap acting on an initial two-photon FH state $\ket{2\,0}$: (a) initial state and FH / SH power $\langle \hat{a}^\dagger_\tau \hat{a}_\tau \rangle$, $\langle \hat{b}^\dagger_\tau \hat{b}_\tau \rangle$ as a function of time. (b) Rabi oscillations visualized in terms of the total SH photon number as well as (c) a Hilbert-space projection onto span($\ket{2\,0}, \ket{0\,1}$) and rotations on the pseudo-Bloch sphere characterized by the pseudo-Pauli operators $\hat{X} = (\hat{a}^{\dagger2} \hat{b} + \hat{a}^2 \hat{b}^{\dagger})/\sqrt{2}$, $\hat{Y} = (\hat{a}^{\dagger2} \hat{b} - \hat{a}^2 \hat{b}^{\dagger})/\sqrt{2}i$, 
$\hat{Z} = \tfrac12 \hat{a}^{\dagger2}\hat{a}^2 - \hat{b}^\dagger \hat{b}$ (these project onto Pauli matrices in the two-state subspace). Here we subtract trivial phase rotations induced by the linear dynamics; see Supplement 1. (d) Gate error of a CZ gate acting on a reference state $\frac{1}{2}(\ket{\tilde{0}}+\ket{\tilde{1}})_1\otimes (\ket{\tilde{0}}+\ket{\tilde{1}})_2$ as a function of the energy gap $\Delta/g$, where subscripts represent the index of qubits. Insets show deviation of output field from the target (input) FH waveform. For all the simulations, we use $\beta_{2,a}=\beta_{2,b}/2$ and $\omega_{b,0}-2\omega_{a,0}=0$ with a large enough system size $T$ to avoid boundary effects. See Supplement 1 for full discussions on numerical simulations.}
\label{fig:f2}
\end{center}
\end{figure*}

This paper studies the third approach: nonlinear enhancement with trapped pulses. The approach is shown in Fig.~\ref{fig:f1}(e), where in addition to the resonant FH and SH fields, we introduce a non-resonant ``trap'' field, generated by an external pulse train, which forms a temporal potential for the resonant, quantum modes. The Hamiltonian for this system takes the form~\cite{Quesada2021}
\begin{align}
	\hat{H} & = \underbrace{\frac{r}{2} \int{\d\tau \left(\hat{a}_\tau^{\dagger 2}\hat{b}_\tau+\hat{a}_\tau^2\hat{b}_\tau^\dagger\right)}}_{\hat{H}_{\text{NL}}} + \sum_{u\in\{a, b\}} \underbrace{\int{\d\tau\,\hat{u}_\tau^\dagger G_u(\tau)\hat{u}_\tau}}_{\hat{H}_{a,\text{L}},\,\hat{H}_{b,\text{L}}}, \label{eq:ham}
\end{align}
with periodic boundary conditions on $-T/2\leq\tau\leq T/2$, where $T$ is the cavity round-trip time (see Supplement 1).

Here, $\hat{a}_\tau$ and $\hat{b}_\tau$ are, respectively, FH and SH field operators with commutation relations $[\hat{a}_\tau,\hat{a}_{\tau'}^\dagger]=[\hat{b}_\tau,\hat{b}_{\tau'}^\dagger]=\delta(\tau-\tau')$, defined in terms of the fast-time coordinate $\tau$~\cite{Lugiato1987} in a co-propagating frame synchronous with the trap field.  $\hat{H}_{\text{NL}}$ represents the $\chi^{(2)}$ interaction, while $\hat{H}_{a,\text{L}}$ and $\hat{H}_{b,\text{L}}$ are the respective linear terms for the FH and SH. For the latter, $G_u(\tau)=D_u(-\mathrm{i}\partial_\tau)+V_u(\tau)$ is a function of the dispersion operator $D_u$ and the trap potential $V_u$ with $u\in\{a,b\}$. The $\chi^{(2)}$ nonlinear coupling constant $r=v_\text{g}\sqrt{\hbar\omega_{b,0}\eta_0}$ is related to group velocity $v_\text{g}$, SH frequency $\omega_{b,0}$, and normalized second harmonic generation (SHG) efficiency $\eta_0$ with units $[\mathrm{power}^{-1}\cdot\mathrm{length}^{-2}]$. As the trapping potential is mediated by XPM, the shape of the temporal trap $V_u(\tau) = -(n_2/n) \omega_u |c_\tau|^2/A$ is determined by the signal frequency $\omega_u$, the trap-pulse power $|c_\tau|^2$, the nonlinear index $n_2$, and the mode area $A$.  Taking into account dispersion up to second order and assuming group-velocity matching between FH and SH, $D_u(s) = \omega_{u,0} -\tfrac12 (\beta_{u,2}/\beta_1) s^2$, where the first and second terms represent the carrier frequency and the group-velocity dispersion (GVD), respectively. The eigenstates of $\hat{H}_{u,\text{L}}$ consist of excitations of normal modes $\Psi_{u,m}(\tau)$ governed by competition between the trap-pulse XPM and GVD, and they are found by solving an eigenmode problem:
\beq
	\label{eq:sch}
	\underbrace{\Bigl(\omega_{u,0}+\frac{\beta_{u,2}}{2\beta_1} \partial_\tau^2 + V_u(\tau)\Bigr)}_{G_u(\tau)}\Psi_{u,m}(\tau)= \lambda_{u,m} \Psi_{u,m}(\tau).
\eeq
In the absence of a trap ($V_u(\tau) = 0$), \eqref{eq:sch} admits continuous wave (CW) eigenmodes $\Psi_{u,m}(\tau) \propto e^{2\pi \mathrm{i}m\tau/T}$, i.e., the usual normal modes of a cavity. In a typical nanophotonic cavity with nonvanishing $\beta_{u,2}$ (Fig.~\ref{fig:f1}(f)), large energy gaps ($\propto \beta_{u,2}T^{-2}$) between eigenmodes ensure that the nonlinear dynamics involve only a single FH/SH mode pair~\cite{Lu2020}.  This scenario properly realizes Hamiltonian \eqref{eq:single-mode-chi2}, but with weak coupling strength due to the large mode volume. Conversely, appropriate dispersion engineering to achieve $\beta_{u,2} \approx 0$ (Fig.~\ref{fig:f1}(g)) makes all modes nearly degenerate, allowing the cavity to support ultrashort pulses. However, this modal degeneracy leads to a major problem: although the nonlinear coupling is increased by the pulse confinement, $\hat{H}_{\rm NL}$ is generally all-to-all, as no mechanism imposes a target pulse shape, leading to intrinsically multimode dynamics unsuitable for high-fidelity qubit operations~\cite{Shapiro2007, Shapiro2006}. These limits highlight the trade-offs between gate fidelity and coupling rate in $\chi^{(2)}$ resonators driven by pulses. Resonators with large $\beta_{u,2}$ driven by long pulses may realize high-fidelity gates with low coupling rates, and conversely, resonators with small $\beta_{u,2}$ driven by short pulses may realize large coupling rates at the cost of reduced gate fidelities. The trap potential eliminates these trade-offs between gate fidelity and coupling rate (see Fig.~\ref{fig:f1}(h)): with anomalous dispersion $\beta_{u,2} < 0$, \eqref{eq:sch} admits at least one {\it bound eigenmode} $\Psi_{u,0}$, localized in time and protected by an energy gap $\Delta_u = |\lambda_{u,1} - \lambda_{u,0}|$. As a result, all spurious couplings to higher-order eigenmodes are suppressed as off-resonance (i.e., phase-mismatched), and the single-mode dynamics of \eqref{eq:single-mode-chi2} are recovered, but with a nonlinear coupling boosted by the temporal confinement of $\Psi_{u,0}$.

The importance of single-mode dynamics to high-fidelity gate operation is highlighted in Fig.~\ref{fig:f2}, where we show the propagation of a signal instantiated in a two-photon FH pulse $\ket{2\,0} = 2^{-1/2}(\hat{a}^\dagger)^2 \ket{0}$, where $\hat{a} = \int\d\tau\,{\Psi^*(\tau)\,\hat{a}_\tau}$ is the annihilation operator for mode $\Psi(\tau)$. To illustrate the limitations of the untrapped case, we first implement $\hat{U}_\pi$ using an input Gaussian waveform $\Psi(\tau)$ with $V_u(\tau) = 0$. Here, the pulse width and chirp are chosen to maximize the gate fidelity given a finite gate time (see Supplement 1), but we observe a rapid decay of Rabi oscillations even for such optimized pulse parameters (see Fig.~\ref{fig:f2}(b)). This observed leakage out of the computational subspace is due to the intrinsically multimode structure of the nonlinear polarization, which couples photons into parasitic temporal modes. These results provide evidence that generic quantum nonlinear propagation of a pulse cannot be described by a single-mode model like \eqref{eq:single-mode-chi2}, posing a nontrivial challenge for NLOQC. This problem is often overlooked in the community, with most proposals assuming a single-mode model without discussing on how single-mode interactions are implemented~\cite{Chuang1995, Nemoto2004, Langford2011, Fukui2022}.

Turning on the temporal trap resolves this problem, restoring effective single-mode dynamics. To show this, we consider the case of a soliton trap $V_a(\tau) = V_b(\tau)/2= -(|\beta_{a,2}|/\beta_1\tau_0^2) \sech^2(\tau/\tau_0)$ with width $\tau_0$, which supports a single bound mode $\Psi_{a,0} = \Psi_{b,0}= (2\tau_0)^{-1/2} \sech(\tau/\tau_0)$. Here, the finite energy gap $\Delta_{a}=\Delta_{b}/2= |\beta_{a,2}|/2\beta_1\tau_0^2$ protects the computational subspace spanned by the bound modes from decoherence, acting as a phase mismatch (i.e.\ detuning) that prevents the nonlinear polarization induced by each bound mode from driving continuum modes. For simplicity, we have assumed the dispersion relationships $\beta_{a,2}=\beta_{b,2}/2$ in this work, but departure from this condition does not qualitatively change the results. The $\chi^{(2)}$ interaction between the FH and SH bound modes becomes phase-matched (i.e., resonant) when $\omega_{b,0}-2\omega_{a,0}=0$, which can be achieved, e.g., by temperature tuning. As a result, effectively single-mode physics reproducing \eqref{eq:single-mode-chi2} is realized between the bound FH and SH modes with coupling constant given by 
\begin{align}
    g=\frac{\pi r}{4\sqrt{2\tau_0}},
\end{align}
which scales as $\tau_0^{-1/2}$ (See Supplement 1). In Fig.~\ref{fig:f2}(a) we show the evolution of a two-photon state instantiated in the FH bound mode, where the photons in the trap are well localized and propagate without dispersing apart from an initial transient. In addition, the dynamics of the SH (Fig.~\ref{fig:f2}(b)) exhibit near-complete Rabi oscillations even for a modest trap with $\Delta/g = 1$, where $\Delta=\Delta_a=\Delta_b/2$. These high-contrast oscillations provide strong evidence of effective single-mode dynamics, which can be further quantified as follows.  Ideally, the gate dynamics are confined within the computational subspace spanned by $\ket{2\,0} = 2^{-1/2}(\hat{a}^\dagger)^2 \ket{0}$ and $\ket{0\,1} = \hat{b}^\dagger \ket{0}$, so we can directly project the system evolution onto span($\ket{2\,0},\,\ket{0\,1}$) in Fig.~\ref{fig:f2}(c). The fact that nearly all of the state amplitude remains in the subspace implies that we have realized the desired single-mode dynamics, i.e., a 180$^{\rm o}$ rotation in the Bloch sphere, picking up a $\pi$ phase shift after returning to the initial state $\ket{2\,0}$.

Gate fidelity scales favorably even for moderate trap depths. In Fig.~\ref{fig:f2}(d), we plot the error $\mathcal{E}$ of a CZ gate as a function of the gap, showing a favorable scaling of $\mathcal{E}\propto (\Delta/g)^{-2}$. For a reference input state, we observe that gate operation with fidelity $>99\%$ is possible with $\Delta/g\gtrsim3$. To visualize the nature of the gate errors, we also show the temporal distribution of the photons; for a shallow trap, photons leak out as dispersive waves, which effectively act as decoherence channels, and incomplete conversion leads to residual SH power. Deepening the trap increases the confinement to the bound mode, suppressing these dispersive waves. Further, the interaction time $t_\pi\propto \tau_0^{-1/2}$ required to implement the gate also shortens for larger trap depth.

\section{Dispersion Engineering and Experimental Prospects}
\label{sec:experiments}

\begin{figure}[h!]
\begin{center}
\includegraphics[width=\columnwidth]{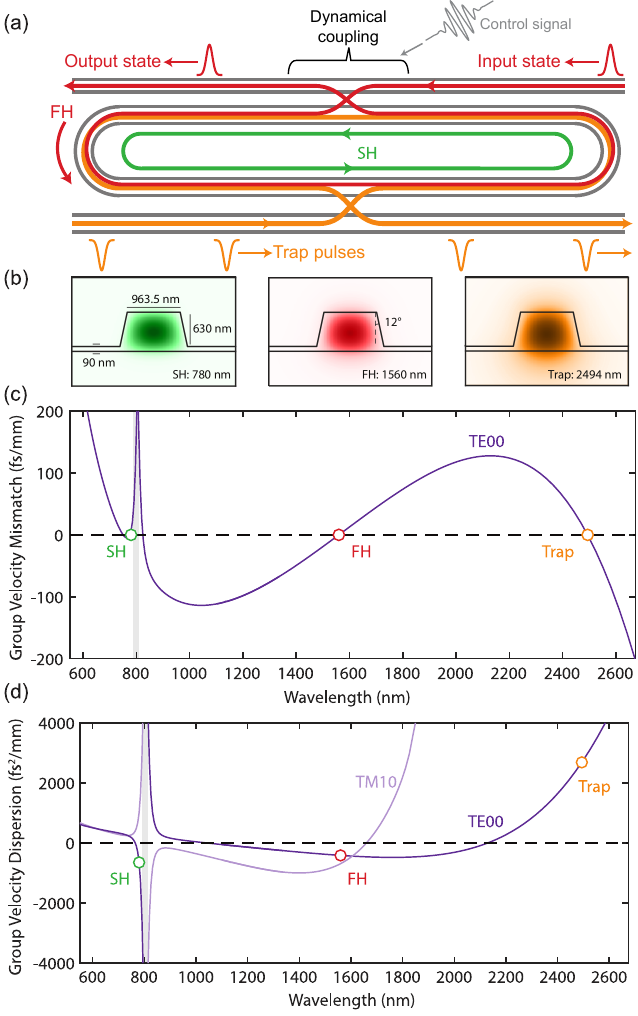}
\caption{Design of a microresonator implementing $\hat{U}_\pi$ with a temporal trap. (a) The FH and trap are coupled into and out of the cavity though two bus waveguides. We assume that the trap pulse is renewed in every round trip, and that quantum input/output states are switched in and out from the cavity by dynamical coupling~\cite{Heuck2020,Heuck2020b,Zhan2019-coupling,Zhu2021}. (b) The waveguide geometry and TE00 field distributions associated with each interacting wave; SH (780 nm), FH (1560 nm) and trap pulse (2494 nm), respectively. (c, d) The group velocity mismatch ($\beta_1 - \beta_{1,a}$) and group velocity dispersion ($\beta_2$) as a function of wavelength. Shaded grey region: avoided crossing between the TE00 and TM10 modes. With a suitable choice of waveguide geometry we may realize both group velocity matching between the FH and SH, and anomalous dispersion for both harmonics. For the ridge geometries considered here, anomalous dispersion may occur at short wavelengths by choosing the location of the avoided crossing to be red-detuned from the SH.}
\label{fig:f3}
\end{center}
\end{figure}

Having established that temporal trapping enables high-fidelity quantum gates with enhanced coupling rates, we now discuss the prospects for experimental realizations in presently available nanophotonics platforms. In realistic situations, photon loss is the primary decoherence channel for quantum gate operations, and to achieve high gate fidelity the nonlinear coupling rate $g$ has to be larger than the characteristic loss rate $\kappa$, which we define as the geometric mean of the FH and SH losses $\kappa = \sqrt{\kappa_a \kappa_{b}}$.  This choice is motivated by analogy to the cooperativity $C \propto g^2/\kappa_\text{cavity}\gamma_\text{atom}$ in cavity QED systems \cite{CarmichaelBook}.

For a ring resonator, the nonlinear coupling between the nominal CW modes is
\begin{align}
\label{eq:cwcoupling}
    g_\text{cw}=\frac{v_\text{g}\sqrt{\hbar\omega_{b,0}\eta_0}}{\sqrt{T}},
\end{align}
where we have used $r=v_\text{g}\sqrt{\hbar\omega_{b,0}\eta_0}$. The round-trip length of the resonator is given by $L=v_\text{g}T$, and a smaller $L$ enhances $g_\text{cw}$ via tighter modal confinement. While microring resonators with radius $\lesssim\SI{100}{\micro m}$ have been realized, bending losses make it challenging to significantly reduce the mode volume further, limiting $g_\text{cw}$ to the order of few megahertz. The same limitation exists for whispering-gallery-mode resonators (WGMRs).  While PhC cavities can realize much smaller wavelength-scale modal confinement and thus a stronger coupling, it is challenging to realize high-Q resonances spanning over an octave, which compromises the overall loss $\kappa$ and results in $g/\kappa$ similar in order of magnitude to ring resonators.

In this context, our prescription allows us to circumvent this trade-off between the mode-volume and the loss: the temporal trap forms a smaller ``flying cavity'' inside a ring resonator, which confines the light further in the axial (temporal) dimension, so that nonlinear interactions between photons benefit from both small mode volume and low loss. Specifically, the nonlinear coupling of the temporally trapped pulses takes the form 
\begin{align}
    g_\text{trap}=\frac{\pi v_\text{g}\sqrt{\hbar\omega_{b,0}\eta_0 }}{4\sqrt{2\tau_0}} = \frac{\pi}{4\sqrt{2}} \sqrt{T/\tau_0}\, g_{\rm cw},
\end{align}
where the width of the trap $\tau_0$ plays the role of the size of an effective cavity. Comparing $g_\text{trap}$ to the CW coupling rate of the same resonator, we find that the coupling is enhanced by the factor proportional to the square root of the pulse duty cycle. Because $g_\text{trap}$ is independent of $T$, temporal trapping may realize large coupling rates for resonators of arbitrary length.

For concreteness, Fig.~\ref{fig:f3} shows a design of a TFLN resonator optimized for implementing our scheme. To couple the quantum states in and out of the resonator with high efficiency, we assume that the coupling between the resonator and the bus waveguide is dynamically controlled, e.g., via nonlinear optical processes~\cite{Brecht2015,Brecht2011,Eckstein2011}. There exist multiple possible implementations of dynamical coupling~\cite{Heuck2020,Heuck2020b,Zhan2019-coupling,Zhu2021} (potentially with their own geometrical constraints and loss considerations), so we keep the following discussions independent of the specific realization. The resonator simultaneously supports a group-velocity matched FH ($\lambda_a=\SI{1560}{nm}$), SH ($\lambda_b=\SI{780}{nm}$), and trap pulse ($\lambda_\text{trap}=\SI{2494}{nm}$). The GVD of both of the harmonics are designed to be anomalous, supporting localized bound modes using bright-pulse XPM. The minimum trap width $\tau_0$ is limited by the dispersion of the trap pulse, for which we assume $\tau_0=\SI{100}{fs}$ to ensure the pulse waveform does not disperse over the propagation through the trapping region. With an estimated SHG efficiency of $\eta_0=\SI{40}{W^{-1}cm^{-2}}$, we obtain a coupling rate of $g_\text{trap}/2\pi=\SI{11.7}{MHz}$. For a 2~mm ring cavity ($T \approx 15$~ps), this is an order larger than the corresponding $g_{\rm cw}$ obtained without trapping.  Moreover, the energy gap of $\Delta/g\approx\text{40}$ provides sufficient isolation of the trapped modes from the continuum. Regarding the loss, $\alpha = 0.7~\text{m}^{-1}$ [$3~\text{dB/m}$] has been achieved in TFLN \cite{Zhang2017}, which through the relation $\kappa = \alpha v_\text{g}$ corresponds to $\kappa/2\pi=\SI{14.4}{MHz}$. These numbers highlight the potential to reach a near-strong-coupling regime $g/\kappa\sim 1$ using ultrashort pulses with technologies available at present. Note that temporal trapping has allowed us to employ a reasonably large resonator size that minimizes the bending loss and sidewall roughness loss, which we expect to make it easier to achieve the loss figure assumed above. Even with propagation loss of $\SI{30}{dB/m}$ (corresponding to $g_\text{cw}/\kappa\sim0.01$), we can achieve $g/\kappa \sim 0.1$.

\begin{figure}[tb]
    \centering
    \includegraphics[width=0.48\textwidth]{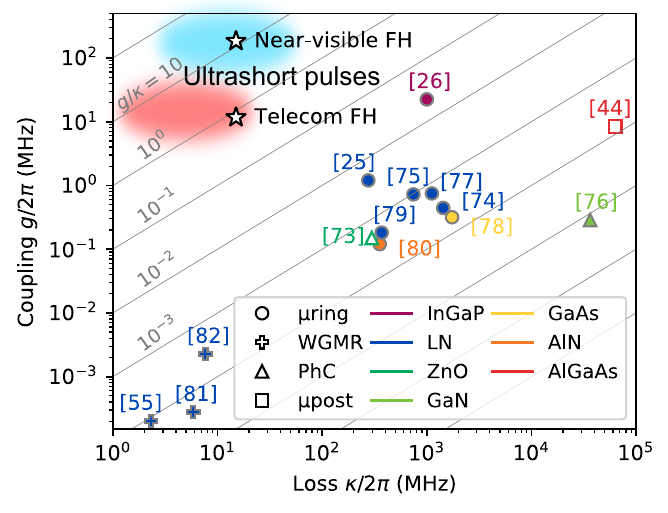}
    \caption{Figure of merit $g/\kappa$ shown for various material platforms and geometries, where the filled and the unfilled markers represent experimental and theoretical results, respectively~\cite{Medina-Vazquez2022,Gao2022,Chen2021,Chen2021-photonconversion,Lu2020,Wang2020,ma2020,Chang2019-GaAs,Chen2019,Bruch2018,Lin2016,Fuerst2010-shg,Fuerst2010-opo,Zhao2022}. When $g$ is not explicitly characterized, we use experimental measures of $\chi^{(2)}$ nonlinearity, e.g., SHG conversion efficiency, to estimate the coupling (see Supplement 1 for full discussions and references). We assume critical coupling, phase-matching between the harmonics, and $\kappa_a=\kappa_b/2$ when the corresponding information is not provided. Stars represent numbers
    estimated for temporally trapped ultrashort pulses at the telecom ($\lambda_a=\SI{1560}{nm}$) and the near-visible ($\lambda_a=\SI{800}{nm}$) FH wavelength.
    }
    \label{fig:f4}
\end{figure}

Further improvements to $g/\kappa$ may be possible in next-generation devices by leveraging the scaling of $g$ with both $\omega$ and $\tau_0$, and by improvements to fabrication processes to reach the material-limited loss rates for $\kappa$. Reductions of the GVD associated with the FH, SH, and trapping pulse enable corresponding reductions of trap pulse duration $\tau_0$, thereby enhancing $g_\mathrm{trap}$. Ultimately, few-cycle operation ($\tau_0 \approx 4\pi/\omega_\mathrm{trap}$) may be made possible with new approaches to dispersion engineering that reduce the GVD of the trapping pulse. Short-wavelength operation increases $g_\mathrm{trap}$ both through the explicit $\omega_{b,0}^{1/2}$ scaling of $g_\mathrm{CW}$ and the $\eta_0\sim \omega^4$~\cite{Jankowski2021-review} scaling associated with the tighter transverse confinement attainable at shorter wavelengths. Recent demonstrations include $\eta_0=\SI{330}{W^{-1}cm^{-2}}$ in a TFLN waveguide at $\lambda_b=\SI{456.5}{nm}$~\cite{Park2022}, and in principle devices with $\eta_0>\SI{1000}{W^{-1}cm^{-2}}$ are possible for FH pulses centered around Ti:sapphire wavelengths~\cite{Jankowski2021-review}. Moreover, the pulse width can be made shorter with a shorter wavelength, i.e., $\tau_0\sim\omega^{-1}$, amounting to a favorable scaling of $g_\text{trap}\sim\omega^3$. Assuming $\lambda_b=\SI{400}{nm}$, which we choose to be below the Urbach tail associated with the material bandgap~\cite{Bhatt2011}, these scalings anticipate the possibility to achieve $g_\text{trap}/2\pi=\SI{184}{MHz}$, corresponding to $g/\kappa>10$. 

In addition, process improvements may reduce losses from the present 3--6~dB/m \cite{Desiatov2019, Zhang2017} by more than an order of magnitude \cite{Shams2021, Gao2021, Gao2022}, limited primarily by bulk material absorption \cite{Ilchenko2004, Serkland1994, Savchenkov2004, Schwesyg2010, Leidinger2015}. For known absorption-limited losses of 0.01 m$^{-1}$, 0.04 m$^{-1}$, and 1 m$^{-1}$ at 1600, 800, and 400 nm, respectively~\cite{Leidinger2015}, we find $\kappa/2\pi = 0.4$ MHz for SHG of telecom photons, and $\kappa/2\pi = 4$ MHz for SHG of 800-nm photons. The large coupling rates made possible by temporal trapping, when combined with absorption-limited losses, provide a pathway to
$g/\kappa > 30$ at telecom wavelengths and $g/\kappa > 40$ at visible wavelengths. We compare these numbers against the current state of the art in a variety of material systems and waveguide geometries in Fig.~\ref{fig:f4}. To date, the highest recorded $g/\kappa$ based on optical nonlinearities is $g/\kappa\approx 10^{-2}$ in a 1560-nm pumped TFLN microresonator~\cite{Lu2020}. In principle, short-wavelength operation and reductions in resonator loss may push conventional CW-pumped nonlinear devices toward $g/\kappa = 0.1$--$1$. In contrast, the $g/\kappa$ enabled by nonlinear resonators using temporal trapping may exceed these limits by two orders of magnitude.

\section{Conclusion}
In this work, we show that temporal trapping can realize strong photon-photon coupling by simultaneously leveraging both temporal and spatial field confinement. The energy gap created between the trapped mode and the remaining cavity modes suppresses undesired multimode interactions, realizing effective single-mode dynamics necessary for high-fidelity quantum gate operations. Our full-quantum simulations confirm that coherent nonlinear dynamics of temporally trapped ultrashort pulses can realize high-fidelity two-qubit entangling gates in a deterministic manner.  This resolves the longstanding concern first raised by Shapiro that pulsed nonlinear optics cannot implement high-fidelity quantum gates \cite{Shapiro2006, Shapiro2007}.

Temporal trapping significantly brightens the prospects of achieving strong coupling in existing photonic platforms~\cite{Irvine2006, Majumdar2013}.  By reducing the effective cavity volume by the pulse duty cycle, $g/\kappa$ can be increased by over an order of magnitude.  Notably, numerical modeling based on realistic dispersion-engineered waveguide designs shows that $g/\kappa\sim1$ is possible on existing TFLN platforms, and true strong coupling $g/\kappa \gg 1$ is plausible with realistic assumptions on wavelength scaling and loss, proposing a unique route towards deterministic optical quantum computation using ultrashort pulses.

Our generic prescription of using temporal trapping to realize enhanced single-mode nonlinear coupling can, in principle, be applied to a broad range of scenarios beyond discrete-variable NLOQC.  For example, continuous-variable implementations of optical quantum computing~\cite{Menicucci2006,Takeda2019} suffer from the same tradeoff between linearity and determinism.  Applied to these systems, strong photon-photon coupling can enable deterministic non-Gaussian gate operations and resource state preparations~\cite{Yanagimoto2020,Zheng2021}, circumventing the need for probabilistic implementations using measurement and feedback. Combined with the ability to manipulate temporal mode structures with optical pulse gating~\cite{Brecht2015,Brecht2011}, deterministic quantum operations on arbitrary photon temporal modes could be realized. Our scheme is compatible with intra-cavity time-multiplexing~\cite{Inagaki2016,Takeda2017} and traveling-wave implementations, enabling unprecedented scalability, qubit uniformity, and operation bandwidth. We expect our work to shed light on the potential to harness ultrafast pulse dynamics for coherent quantum computation and engineering, guiding ongoing experimental and theoretical efforts towards this unique frontier of broadband quantum optics.

\section*{Funding}
Army Research Office (W911NF-16-1-0086); National Science Foundation (CCF-1918549, PHY-2011363).
\section*{Acknowledgements}
The authors wish to thank NTT Research for their financial and technical support. R.\,Y. is supported by a Stanford Q-FARM Ph.D. Fellowship and the Masason Foundation.
\section*{Disclosures}
RY, EN, MJ, RH (P).

\section*{Data availability}
Data underlying the results presented in this paper are not publicly available at this time but may be obtained from the authors upon reasonable request.
\section*{Supplemental document}
See Supplement 1 for supporting content.

\newpage

\setcounter{equation}{0}
\setcounter{figure}{0}
\setcounter{section}{0}
\setcounter{table}{0}
\makeatletter
\renewcommand{\thetable}{S\arabic{table}}
\renewcommand{\thefigure}{S\arabic{figure}}
\renewcommand{\thesection}{S\arabic{section}}
\renewcommand{\thesubsection}{S\arabic{section}\Alph{subsection}}
\renewcommand{\theequation}{S\arabic{equation}}
\renewcommand{\thefigure}{S\arabic{figure}}
\renewcommand{\bibnumfmt}[1]{[S#1]}
\renewcommand{\citenumfont}[1]{S#1}

\section*{[Supplementary Material] Temporal trapping: a route to strong coupling and deterministic optical quantum computation}

R.~Yanagimoto$^{1,*}$, E.~Ng$^{1,2}$, M.~Jankowski$^{1,2}$, H.~Mabuchi$^{1}$, and R.~Hamerly$^{2,3,\dagger}$

\noindent $^1${\it E.\,L.\,Ginzton Laboratory, Stanford University, Stanford, California 94305, USA} \\
$^2${\it Physics \& Informatics Laboratories, NTT Research, Inc., Sunnyvale, California 94085, USA} \\
$^3${\it Research Laboratory of Electronics, MIT, 50 Vassar Street, Cambridge, MA 02139, USA} \\
$^*${ryotatsu@stanford.edu} \\
$^\dagger${rhamerly@mit.edu}

\noindent \rule{\linewidth}{0.5pt}

\section{Nonlinear Hamiltonian}
\label{sec:a1}

\subsection{Generic Form}

This section derives the general Hamiltonian for a nonlinear-optical cavity.  A fully rigorous derivation is very involved, as one must account for dispersion in the linear and nonlinear polarizabilities, and take care to properly quantize fluctuations using the $D$ field \cite{S_Drummond1990, S_Raymer2020, S_Quesada2017}.  However, for weakly dispersive materials and perturbative nonlinearities, a simpler phenomenological model suffices \cite{S_Rodriguez2007}.  To start, assume a wavelength-independent refractive index.  We write the electric field as a sum of normal-mode fluctuations
\beq
	E(\vec{x}, t) = \sum_m \sqrt{\hbar\omega_m/2\epsilon_0} \bigl(A_m(t) E_m(\vec{x}) + \text{c.c.}\bigr), \label{eq:equant}
\eeq
where the $E_m(\vec{x})$ are normalized to $\int{n(\vec{x})^2 |E_m(\vec{x})|^2 \d^3 x} = 1$.  We first solve Eq.~(\ref{eq:equant}) by treating $A_m(t)$ as classical variables, which must evolve to satisfy the Helmholtz equation:
\beq
	\nabla\times (\nabla\times E) + \frac{n^2}{c^2} \frac{\partial^2 E}{\partial t^2} = -\frac{1}{c^2} \frac{\partial^2(\delta P/\epsilon_0)}{\partial t^2} \label{eq:hhz}
\eeq
In the absence of a perturbing term $\delta P$, $\dot{A}_m = -i\omega_m A_m$, since the $E_m$ are normal modes.  Linear perturbations $\delta P/\epsilon_0 = \delta\epsilon_r E$ alter the mode frequencies via the well-known expression $\delta\omega_m = -\tfrac12 \omega_m \int{\delta\epsilon_r |E_m|^2 \d^3\vec{x}}$ \cite{S_JoannopoulosBook}.  This expression can be generalized to \cite{S_Rodriguez2007}:
\beq
    \delta\dot{A}_m = \frac{i\omega_m}{2\sqrt{\hbar\omega_m/2\epsilon_0}} \int{E_m^* \delta P\, \d^3\vec{x}} \label{eq:adot_gen}
\eeq

\subsection{Parametric ($\chi^{(2)}$) Material}
\label{sec:app-para}

For the parametric nonlinearity, $\delta P_i = \tfrac12 d_{ijk} E_j E_k \equiv \tfrac12 d: EE$.  The resulting equation of motion, simplified using Kleinman symmetry \cite{S_Boyd2008} is:
\begin{align}
    \dot{A}_m & = -i\omega_m A_m + i \sum_{np} \sqrt{\frac{\hbar\omega_m\omega_n\omega_p}{2\epsilon_0}} \Bigl[A_n A_p \int_*{d:E_m^* E_n E_p \d^3\vec{x}} \nonumber \\
    & \qquad\qquad\qquad\qquad\qquad\qquad\qquad + 2 A_n^* A_p \int_*{d:E_m^* E_n^* E_p \d^3\vec{x}}\Bigr] \nonumber \\
    & = -i\omega_m A_m + \frac{i}{2}\sum_{np} \bigl(g_{m,np} A_n A_p + g_{p,mn}^* A_n^* A_p\bigr),
    \nonumber \\
    & \qquad \Bigl(g_{m,np} \equiv \sqrt{\frac{2\hbar\omega_m\omega_n\omega_p}{\epsilon_0}} \int_*{d:E_m^* E_n E_p \d^3\vec{x}}\Bigr) \label{eq:adot}
\end{align}
(Here $\int_*(\ldots)\d^3\vec{x}$ refers to an integral restricted to the nonlinear material.)  Now we quantize the fields, replacing $A_m \rightarrow \hat{A}_m$ with canonical commutators $[\hat{A}_m, \hat{A}_n] = 0$, $[\hat{A}_m, \hat{A}_n^\dagger] = \delta_{mn}$.  Eq.~(\ref{eq:adot}) is generated by the following Hamiltonian:
\beq
    \hat{H} = \sum_m {\omega_m \hat{A}_m^\dagger \hat{A}_m} 
    - \frac{1}{2} \sum_{mnp} \bigl(g_{m,np} \hat{A}_m^\dagger \hat{A}_n \hat{A}_p + \text{h.c.}\bigr) \label{eq:hamgen}
\eeq
The simplest case involves a two-mode cavity where FH $\hat{a}$ and SH $\hat{b}$.  Here, up to a phase, $\hat{H}_{\rm NL} = \tfrac12 g \bigl(a^2 b^\dagger + (a^\dagger)^2 b\bigr)$, where $g$ is given by:
\beq
    g \equiv g_{b,aa} = \sqrt{\frac{4\hbar\omega^3}{\epsilon_0}} \int_*{d:E_b^* E_a E_a \d^3\vec{x}}
\eeq
Here, as before, we have energy-normalized the modes as $\int{\epsilon_r |E|^2 \d\vec{x}} = 1$.

Two related quantities are often used to quantify the nonlinear interaction: the effective mode volume $V_{\rm sh}$ (usually normalized as $\tilde{V}_{\rm sh} = V_{\rm sh}/(\lambda/n)^3$ and the nonlinear overlap $\bar{\beta}$:
\begin{align}
    \tilde{V}_{\rm sh} & = \Bigl| n^3 \int_*{\hat{d}: E_b^* E_a E_a \d^3\vec{x}} \Bigr|^{-2},\nonumber \\
    \bar{\beta} & = \lambda^{3/2} \int_*{\hat{d}: E_b^* E_a E_a \d^3\vec{x}}
    = \frac{1}{\sqrt{n^3 \tilde{V}_{sh}}} \label{eq:vsh}
\end{align}
Here, $\hat{d}_{ijk} = d_{ijk}/d_{\rm eff}$ is the normalized nonlinear tensor.  The volume is defined relative to an ``ideal'' cavity: for hypothetical flat-top modes with constant norm $|E_a|, |E_b| = \text{const}$ and perfect phase-matching, $V_{\rm sh}$ will return the physical cavity volume.  Note, however, that poor mode overlap can cause $\tilde{V}_{\rm sh}$ to be much larger than the actual volume of either mode.

Relative to these quantities, $g$ is given by:
\beq
    g = \frac{4 d_{\rm eff}}{\lambda^3} \sqrt{\frac{2 \pi^3 \hbar c^3}{n^3 \epsilon_0 \tilde{V}_{\rm sh}}} = \frac{4 d_{\rm eff} \bar{\beta}}{\lambda^3} \sqrt{\frac{2 \pi^3 \hbar c^3}{\epsilon_0}}
\eeq
The effective loss rate is $\kappa \equiv \sqrt{\kappa_a \kappa_b} = (2\pi c/\lambda) \sqrt{2/Q_a Q_{b}}$.  Dividing these quantities yields the expression for $g/\kappa$, Eq.~(3) in the main text.  For LiNbO$_3$ at $\lambda = 1.55~\mu{\rm m}$ ($n = 2.2$, $d_{33} = 33~\text{pm/V}$), this expression yields $g/\kappa \approx 10^{-6} \sqrt{Q_a Q_{b}/\tilde{V}_{\rm sh}}$.  This figure of merit is calculated for representative cavities in  Table~\ref{tab:t0-2}.  Despite the much larger mode volume, a doubly-resonant ring cavity is expected to have a higher $g/\kappa$ due to the larger $Q_b$, and temporal trapping improves the figure still further by reducing the effective volume.

\subsection{Kerr ($\chi^{(3)}$) Material}
\label{sec:kerr}


\begin{table}[tb]
\begin{center}
\begin{tabular}{c|cc|c|c|c}
\hline\hline
& \ \ $Q_{a}$\ \ & \ \ $Q_{b}$\ \  &
\ \ $\tilde{V}_{\rm sh}$\ \ \  & $d_{\rm eff}/d_{33}$ & 
$C_{\chi^{(2)}}$  \\ \hline
PhC$^{*}$ & $10^6$ & $10^3$ & $1$ & $1$ & 0.03 \\
Ring$^\dagger$ & $10^7$ & $10^7$ & $2000$ & $2/\pi$ & 0.1 \\
Trapped$^\ddag$ & $10^7$ & $10^7$ & $40$ & $2/\pi$ & 0.8 \\  \hline\hline
\end{tabular}
\end{center}
\caption{Estimates of $Q$, $V$, and cooperativity $C_{\chi^{(2)}} = g/\kappa$ for a quadratic nonlinearity under competing confinement mechanisms.  LiNbO$_3$, $\lambda = 1.55$~$\mu$m.  $^*$PhC cavities typically have $\tilde{V} \sim 1$.  Tip cavities can in principle achieve deep-subwavelength volume with respect to atom-cavity coupling and the Kerr effect \cite{S_Hu2016, S_Choi2017}, but it is difficult to make $\tilde{V}_{\rm sh} \ll 1$ due to the weak divergence of the integrand in Eq.~(\ref{eq:vsh}).  Note that $\tilde{V}_{\rm sh} \sim 1$ is a lower estimate, as poor mode overlap can greatly increase the effective volume.  $^\dagger$Ring circumference 2~mm, loss $\alpha = 3$~dB/m \cite{S_Zhang2017}.  $^\ddag$Pulse of width 100~fs, loss $\alpha = 3$~dB/m. 
}
\label{tab:t0-2}
\end{table}

For comparison, we also provide estimates for the nonlinear coupling in a Kerr material.  Here, the nonlinear polarization $P_{\rm NL}/\epsilon_0 = \chi: EEE$ is cubic in $E$.  Applying \eqref{eq:adot_gen} and ignoring the off-resonant third-harmonic terms, we find:
\begin{align}
    \dot{A}_m & = -i\omega_m A_m + i \sum_{npq} \chi_{mn,pq} A_n^* A_p A_q,\nonumber \\
    \chi_{mn,pq} & = \frac{3\hbar \sqrt{\omega_m\omega_n\omega_p\omega_q}}{4\epsilon_0} \int_*{\chi:E_m^*E_n^*E_pE_q \d^3\vec{x}}
\end{align}
Upon quantizing the fields, this corresponds to the Hamiltonian:
\beq
    \hat{H} = \sum_m{\omega_m \hat{A}_m^\dagger\hat{A}_m} - \frac{1}{2} \sum_{mn,pq}{\chi_{mn,pq} \hat{A}_m^\dagger\hat{A}_n^\dagger\hat{A}_p\hat{A}_q}
\eeq
Reducing this to a singly-resonant cavity with field $\hat{a}$, we obtain $\hat{H} = \omega \hat{a}^\dagger \hat{a} - \tfrac12 \chi \hat{a}^\dagger\hat{a}^\dagger\hat{a}\hat{a}$, where
\begin{align}
    \chi & = -\frac{3\hbar \omega^2}{4\epsilon_0} \int_*{\chi:E^*E^*EE\, \d^3\vec{x}}
    = \frac{3\pi^2 \hbar c^2 \chi_{\rm eff}}{n \epsilon_0 \lambda^5 \tilde{V}_{rm k}}, \nonumber \\
    V_k & = \Bigl(n^4 \int_*{\hat{\chi}: E^*E^*EE\, \d^3\vec{x}}\Bigr)^{-1} \label{eq:vk}
\end{align}
Here, as before, $\hat{\chi} = \chi/\chi_{\rm eff}$ is the normalized Kerr tensor.  Given $\kappa = 2\pi c/\lambda Q$, the figure or merit for the Kerr cavity is:
\beq
    \frac{\chi}{\kappa} = \frac{3\pi \hbar c \chi_{\rm eff}}{2n \epsilon_0 \lambda^4} \frac{Q_a}{\tilde{V}_{\rm k}}
\eeq
Direct-gap III-V semiconductors such as AlGaAs and InGaP are promising platforms for $\chi^{(3)}$ nonlinear optics given their relatively high nonlinear index, large index contrast permitting tight bending radii, and lack of two-photon absorption at telecom wavelengths \cite{S_pu2016efficient}.  For AlGaAs at $1.55~\mu{\rm m}$ ($n = 3.3$, $\chi = 0.8~\text{nm}^2/\text{V}^2$ \cite{S_wathen2014efficient, S_lacava2014nonlinear, S_dolgaleva2010broadband}), $\chi/\kappa \approx (7\times 10^{-10}) Q_a/\tilde{V}_{\rm k}$.  Table~\ref{tab:t0-2} compares the cooperativity of $\chi^{(2)}$ and $\chi^{(3)}$ mechanisms for the PhC, ring, and trapped-pulse situations.  

Two factors favor wavelength-scale resonators in the Kerr case.  First, only a single resonance is required, and singly-resonant PhC cavities can have very large $Q$ factors.  Second, the integrand in Eq.~(\ref{eq:vk}) is proportional to $|E|^4$ (as opposed to $|E|^3$ in Eq.~(\ref{eq:vsh})); this means that in tip-cavity structures with divergent field profiles \cite{S_Hu2016, S_Choi2017}, the integrand diverges more strongly, allowing smaller effective mode volumes.  This, combined with the stronger volume dependence $\chi/\kappa \sim V_{\rm k}^{-1}$, significantly favors PhC cavities, even though rings can be made smaller in $\chi^{(3)}$ platforms such as silicon- or AlGaAs-on-insulator due to the higher index contrast.  Despite these advantages, it is challenging to see a scenario in which the single-photon anharmonicity can be increased beyond 0.1, suggesting that doubly-resonant quadratic nonlinearities (or induced $\chi^{(2)}$, e.g.\ via electric \cite{S_Timurdogan2017} or optical \cite{S_Langford2011} fields) are a more promising route to all-optical strong coupling.

\begin{table}[tb]
\begin{center}
\begin{tabular}{c|c|c|c}
\hline\hline
& \ \ $Q_{a}$\ \ &
\ \ \ $\tilde{V}_{\rm k}$\ \ & $C_{\chi^{(3)}}$  \\ \hline
PhC$^{*}$ & $10^6$ & $10^{-2}$ & 0.07 \\
Ring$^\dagger$ & $10^7$ & $100$ & $7 \times 10^{-5}$ \\
Trapped$^\ddag$ & $10^7$ & $40$ & $2 \times 10^{-4}$ \\  \hline\hline
\end{tabular}
\end{center}
\caption{Estimates of $Q$, $V$, and cooperativity $C_{\chi^{(3)}} = \chi/\kappa$) for a Kerr nonlinearity under competing confinement mechanisms. AlGaAs, $\lambda = 1.55\mu$m.  $^*$Tip-cavity PhC engineered to have a deep-subwavelength volume \cite{S_Hu2016, S_Choi2017}.  $^\dagger$Ring circumference 100~$\mu$m, loss $\alpha = 3$~dB/m \cite{S_Zhang2017}.  $^\ddag$Pulse of width 100~fs, loss $\alpha = 3$~dB/m. 
}
\label{tab:t0-3}
\end{table}

\section{\uppercase{Ring Cavity and Normalization}}
\label{sec:a2}

\begin{figure*}[t!]
\begin{center}
\includegraphics[width=0.75\textwidth]{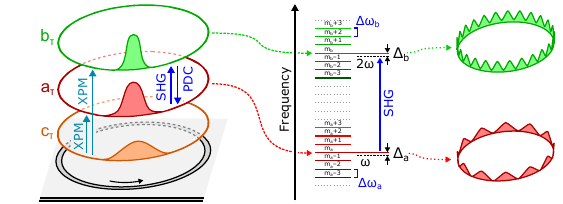}
\caption{Pulses propagating along the ring resonator in the two-band model.  Spectrum is divided into FH $\hat{a}_m$ and SH $\hat{b}_m$ modes, while the trap field is nonresonant.}
\label{fig:fa1}
\end{center}
\end{figure*}

\subsection{Ring Cavity}

Ring cavities enjoy cylindrical symmetry, so the eigenmodes, which can be found with separation of variables, integer angular quantum number (i.e. dependence $E_m \sim e^{im\phi}$).  Most rings are large enough that they can be modeled as waveguides that ``wrap around'' at length $L = 2\pi R$; in this case, the axial variable is $z \equiv R\phi$ and the transverse variables are $x \equiv \rho-R$ and $y$.  The modes $E_{m,n}(\vec{x}) = L^{-1/2} E_{m,n}^{\perp}(x, y) e^{2\pi imz/L}$ are indexed by axial and transverse quantum numbers $(m,n)$.  Due to phase-matching conditions, only a single transverse mode participates meaningfully in the dynamics, so to simplify the notation, we drop the transverse index $n$ and the superscript $\perp$ when writing $E$: $E_{m,n}^\perp(x, y) \rightarrow E_m(x, y)$.

The Hamiltonian takes the form of \eqref{eq:hamgen}, where the coupling elements are given by:
\beq
    g_{m,np} = \sqrt{\frac{2\hbar\omega_m \omega_n \omega_p}{\epsilon_0 L}} \tilde{\pi}_{m-n-p} \int_*{d: E_m^* E_n E_p \d A} \label{eq:g-ring}
\eeq
where the $\tilde{\pi}_m$ are the Fourier series coefficients of the poling function $\pi(z): [0, L] \rightarrow \{-1, +1\}$, defined as $\tilde{\pi}_m = L^{-1} \int{\pi(z) \cos(2\pi m z/L) \d z}$.  For modal phase-matching, $\pi(z) = 1$ and $\tilde{\pi}_m = \delta_{m0}$.  For quasi-phase matching with 50\% duty cycle, $\pi(z) = \text{sign}(\cos(2\pi s z/L))$, where $s \in \mathbb{Z}$ is the number of poling periods per circumference.  The QPM factor works out to $\tilde{\pi}_{\pm s} = 2/\pi$ (higher-order QPM processes are not relevant here).

Since the trapped pulses comprise many optical cycles, their bandwidth is narrow compared to the carrier frequency.  In this case, we can use a two-band model that separates the FH ($a_m$) and SH ($b_m$), depicted in Fig.~\ref{fig:fa1}:
\beq
	\biggl\{
	\begin{array}{lll}
	\hat{a}_m = \hat{A}_{m_a + m}, & \hat{b}_m = \hat{A}_{m_b + m} \\
	\omega_{a,m} = \omega_{m_a + m}, & \omega_{b,m} = \omega_{m_b + m} 
	\end{array}
\eeq
In the two-band model, the pulses are narrow-band enough that the cross-sectional fields $E_{a,m} \equiv E_{m+m_a}$, $E_{b,m} \equiv E_{m+m_b}$ depend only weakly on index $m$; we can therefore suppress the index $E_{a,m} \rightarrow E_a$, $E_{b,m} \rightarrow E_b$.  The cavity is poled to phase-match $\hat{a}_0$ and $\hat{b}_0$, i.e.\ $s = m_b - 2m_a$.  Under these assumptions, the Hamiltonian Eq.~(\ref{eq:hamgen}) reduces to:
\beq
	H = \sum_m \bigl(\omega_{a,m} \hat{a}_m^\dagger \hat{a}_m + \omega_{b,m} \hat{b}_m^\dagger \hat{b}_m \bigr)
	- \frac{r}{2\sqrt{T}} \sum_{mn} \bigl(\hat{b}_{m+n}^\dagger \hat{a}_m \hat{a}_n + \hat{b}_{m+n} \hat{a}_m^\dagger \hat{a}_n^\dagger\bigr)
\eeq
where $T = n_g L/c$ is the cavity repetition rate at a designated group velocity $c/n_g$ (defined below), and
\beq
    r = 2 d_{\rm eff} \sqrt{\frac{\hbar\omega^3 n_g}{\epsilon_0 c}} \int_*{\hat{d}: E_b^* E_a E_a \d A} = \frac{4 d_{\rm eff}}{n^2} \sqrt{\frac{2 \pi^3 n_g \hbar c^2}{\epsilon_0 \lambda^5 \tilde{A}_{\rm sh}}}
\eeq
is the parametric interaction strength.  As in Sec.~\ref{sec:app-para}, we define an effective mode area $A_{\rm sh} = \bigl| n^3 \int_*{\hat{d}:E_b^* E_a E_a \d A} \bigr|^{-2}$, where $\tilde{A}_{\rm sh} = A_{\rm sh}/(\lambda/n)^2 \sim 1$ for high-index-contrast waveguides.  We can relate $r$ to the normalized SHG efficiency $\eta_0$, a value experimentally quoted for waveguides \cite{S_Parameswaran2002, S_Kurimura2006, S_Wang2018, S_Park2022}, as follows:
\begin{align}
    r & = \sqrt{\frac{4\pi c^3 \hbar \eta_0}{\lambda n_g^2}} = v_g \sqrt{2\hbar\omega \eta_0},\nonumber \\
    \eta & \equiv \frac{8\pi^2 Z_0 n_g^3 d_{\rm eff}^2}{\lambda^2} \Bigl|\int_* \hat{d}: E_b^* E_a E_a \d A \Bigr|^{-2} = \frac{8\pi^2 Z_0 n_{g}^3 d_{\rm eff}^2}{\lambda^4 n^4 \tilde{A}_{\rm sh}}
\end{align}
We perform a rotating-wave transformation to move into the co-propagating basis with designated carrier frequency $\omega$ and repetition rate $\Omega$ (which corresponds to group index $n_g = 2\pi c/L\Omega$): $\hat{a}_m \rightarrow e^{-i(\omega + m\Omega) t} \hat{a}_m$, $\hat{b}_m \rightarrow e^{-i(2\omega + m\Omega) t} \hat{b}_m$.  The effect of this transformation is to shift the energy levels $\omega_{a,m}, \omega_{b,m}$:
\beq
    \omega_{a,m} \rightarrow \omega_{a,m} - (\omega + \Omega t),\ \ \ 
    \omega_{b,m} \rightarrow \omega_{b,m} - (2\omega + \Omega t)
\eeq
In the dispersion-engineered case, the FH and SH pulses travel at nearly-matched group velocities $n_{a,g} \approx n_{b,g} \approx n_g$ and $T$ is chosen as a corresponding average round-trip time.  Specifically, for a given $n_g = c T/L$, define $\Delta n_{u,g} = n_{u,g} - n_g$.  Up to quadratic order in dispersion, the resonance frequencies in the rotating frame are:
\beq
	\omega_{u, m} = \omega_{u,0} - \frac{\Delta n_{u,g}}{n_g^2} \Omega m + \frac{\beta_{u,2} c}{2n_g} \Omega^2 m^2 \label{eq:wrf}
\eeq
Finally, we perform a Fourier transform to convert the the propagating fields $\hat{a}_m$, $\hat{b}_m$ into a temporal basis, depending on the ``fast time'' $\tau$:
\beq
	a(t, \tau) = \frac{1}{\sqrt{T}}\sum_m{a_m(t) e^{2\pi i m \tau/T}},\ \ \ 
	b(t, \tau) = \frac{1}{\sqrt{T}}\sum_m{b_m(t) e^{2\pi i m \tau/T}}
\eeq
These fields are normalized so that $[a(t, \tau), a^\dagger(t, \tau')] = \delta(\tau-\tau')$, etc., and are periodic in $\tau$.  In the group-velocity matched case $\Delta n_{u,g} = 0$, the resulting Hamiltonian is:
\begin{align}
    \hat{H}=\frac{r}{2}\int\mathrm{d}\tau\,\left(\hat{a}_\tau^{\dagger 2}\hat{b}_\tau+\hat{a}^2_\tau\hat{b}^\dagger_\tau\right)+\sum_{u\in\{a,b\}}\int\mathrm{d}\tau\,\hat{u}_\tau^\dagger \left(\omega_{u,0}+\frac{\beta_{u,2}}{2\beta_1}\partial_\tau^2\right)\hat{u}_\tau, 
\end{align}

Finally, we introduce the trapping field.  This field is not resonant with the cavity, so it can be treated as an external $\tau$-dependent potential term in the Hamiltonian.  This yields the final form of the Hamiltonian, which is given in Eq.~(4) in the main text: 
\begin{align}
    \hat{H} & = \frac{r}{2}\int\mathrm{d}\tau\,\left(\hat{a}_\tau^{\dagger 2}\hat{b}_\tau+\hat{a}^2_\tau\hat{b}^\dagger_\tau\right) \nonumber \\
    & \qquad +\ \sum_{u\in\{a,b\}}\int\mathrm{d}\tau\,\hat{u}_\tau^\dagger \left(\omega_{u,0}+\frac{\beta_{u,2}}{2\beta_1}\partial_\tau^2+V_u(\tau)\right)\hat{u}_\tau, \label{eq:dimham}
\end{align}
where $V_b(\tau)=2V_a(\tau)$.

\subsection{Normalization}

In this section, we describe how we transform the dimensionful Hamiltonian (\ref{eq:dimham}) of the $\chi^{(2)}$-nonlinear waveguide into a dimensionless form well suited for numerical simulation. Furthermore, nondimensionalization also allows us to extract essential dimensionless parameters which uniquely characterize the system dynamics. For simplicity, we assume here that both harmonics have equal group velocities and anomalous group velocity dispersions, but the following can readily be extended to handle more general cases.

First, we introduce characteristic fast and slow timescales $t_\text{c}=\left(|\beta_{a,2}|/r^4\beta_1\right)^{1/3}$ and $\tau_\text{c}=\left(|\beta_{a,2}|/r\beta_1\right)^{2/3}$, respectively. This allows us to rewrite the Hamiltonian as
\begin{align}
    \hat{H} & =\frac{1}{t_\text{c}}\biggl\{\frac{1}{2}\int\mathrm{d}\xi\,\left(\hat{a}_\xi^{\dagger2}\hat{b}_\xi+\hat{a}_\xi^{2}\hat{b}_\xi^\dagger\right) \nonumber \\
    & +\int\mathrm{d}\xi \hat{a}_\xi^\dagger\left(-\frac{1}{2}\partial_\xi^2+\omega_{a,0}t_\text{c}+U(\xi)\right)\hat{a}_\xi \nonumber \\
    & +\int\mathrm{d}\xi \hat{b}_\xi^\dagger\left(-\frac{\rho}{2}\partial_\xi^2+\omega_{b,0}t_\text{c}+2U(\xi)\right)\hat{b}_\xi\biggr\}, \label{eq:normalized-hamiltonian}
\end{align}
where $\xi=\tau/\tau_\text{c}$ is the normalized (dimensionless) fast-time coordinate, $\rho=\beta_{b,2}/\beta_{a,2}$ is the ratio between the SH and FH GVDs, and $U(\xi)=V_a(\xi\tau_\text{c})t_\text{c}=\frac{1}{2}V_b(\xi\tau_\text{c})t_\text{c}$ is the normalized potential. Note that these specific choices of $t_\text{c}$ and $\tau_\text{c}$ fix the coefficients for the nonlinear coupling and the FH GVD to a canonical value of $1/2$.

Next, we move to a rotating frame of the FH via the operator mappings $\hat{a}\mapsto e^{-\mathrm{i}\omega_{a,0}t}\hat{a}$ and $\hat{b}\mapsto e^{-2\mathrm{i}\omega_{a,0}t}\hat{b}$, after which we obtain
\begin{align}
\label{eq:rotating-hamiltonian}
\begin{split}
    \hat{H}=\frac{1}{t_\text{c}}\left\{\frac{1}{2}\int\mathrm{d}\xi\,\left(\hat{a}_\xi^{\dagger2}\hat{b}_\xi+\hat{a}_\xi^{2}\hat{b}_\xi^\dagger\right)+\int\mathrm{d}\xi \hat{a}_\xi^\dagger\left(-\frac{1}{2}\partial_\xi^2+U(\xi)\right)\hat{a}_\xi\right.\\\left.+\int\mathrm{d}\xi \hat{b}_\xi^\dagger\left(\delta-\frac{\rho}{2}\partial_\xi^2+2U(\xi)\right)\hat{b}_\xi\right\},
\end{split}
\end{align}
where we have introduced the normalized phase mismatch $\delta=(\omega_{b,0}-2\omega_{a,0})t_\text{c}$. At this point, we note that \eqref{eq:rotating-hamiltonian} has only three dimensionless quantities that nontrivially determine the system dynamics, i.e., $U$, $\delta$, and $\rho$. As a result, two systems with identical $U$, $\delta$, and $\rho$ after normalization exhibit quantum dynamics that are equivalent up to simple rescalings of the two time coordinates.

As a further simplification, we consider in the main text a sech-shaped trap $V_a(\tau)=V_b(\tau)/2=-(|\beta_{a,2}|/\beta_1\tau_0^2)\sech^2(\tau/\tau_0)$ with zero phase-mismatch $\omega_{b,0}-2\omega_{a,0}=0$ and $\beta_{a,2}=\beta_{b,2}/2$, which in dimensionless form corresponds to $U(\xi)=-\xi_0^{-2}\sech^2(\xi/\xi_0)$, $\delta=0$, and $\rho=2$. As a result, the normalized trap width $\xi_0=\tau_0/\tau_{\mathrm{c}}$ is left as the sole parameter which uniquely determines the system dynamics. In particular, $\xi_0$ determines the ratio between the characteristic energy gap $\Delta=\Delta_a$ and the nonlinear coupling between the bound mode $g$ via the reciprocal relationship
\begin{align}
    \frac{\Delta}{g}=\frac{2\sqrt{2}}{\pi}\xi_0^{-3/2},
\end{align}
which indicates that we can also equivalently use $\Delta/g$ to uniquely characterize the system dynamics.

\section{\uppercase{Numerical simulation of the quantum pulse propagation}}
\label{sec:a3}
In general, full quantum simulations of the pulse propagation dynamics can be readily performed by leveraging techniques such as matrix product states~\cite{S_Yanagimoto2021_mps} or supermode expansion~\cite{S_Yanagimoto2021-non-gaussian}, especially when written in the dimensionless form introduced in Sec.~\ref{sec:a2}.

However, in this work, we are primarily concerned with quantum pulses containing only up to two FH or one SH photons, which can be more directly and concisely captured using a wavefunction of the general form
\begin{align}
    \ket{\varphi(t/t_\text{c})} & =\biggl(P+\int\mathrm{d}\xi\,Q_\xi\hat{a}_\xi^\dagger +  \int\mathrm{d}\xi\,S_\xi\hat{b}_\xi^\dagger \nonumber \\
    & \qquad +\int\mathrm{d}\xi_1\mathrm{d}\xi_2\,R_{\xi_1,\xi_2}\hat{a}_{\xi_1}^\dagger\hat{a}_{\xi_2}^\dagger \biggr)\ket{0}. \label{eq:two-photon-state}
\end{align}
The time evolution of the wavefunction under \eqref{eq:rotating-hamiltonian} can then be shown to obey $\frac{\partial P}{\partial (t/t_\text{c})}=0$ and
\begin{align}
    \begin{split}
    \mathrm{i}\frac{\partial Q_\xi}{\partial (t/t_\text{c})}&=\left(-\frac{1}{2}\partial_\xi^2+U(\xi)\right)Q_\xi\\
    \mathrm{i}\frac{\partial S_\xi}{\partial (t/t_\text{c})}&=\left(\delta-\frac{\rho}{2}\partial_\xi^2+2U(\xi)\right)S_\xi+R_{\xi,\xi}(t)\\
    \mathrm{i}\frac{\partial R_{\xi_1,\xi_2}}{\partial (t/t_\text{c})}&=\left(\!-\frac{\partial_{\xi_1}^2+\partial_{\xi_2}^2}{2}+U(\xi_1)+U(\xi_2)\!\right)\!R_{\xi_1,\xi_2}+\frac{1}{2}\delta(\xi_1-\xi_2)S_{\xi_1},
\end{split}
\end{align}
which can be efficiently integrated, e.g., by split-step Fourier methods.

\section{\uppercase{Temporal supermodes for pulsed quantum gate operations}}
\label{sec:a4}
In discussing quantum gate operations with photonic qubits, it is often implicitly assumed that the computational mode of the qubit is static and well defined for all time, and this is often the case in single-mode quantum systems such as microring resonators or photonic crystal cavities. For a multimode pulsed system, however, computational modes can take the form of any collective excitation  $\int\mathrm{d}\tau\,\Psi(\tau)\hat a_\tau$, so a complete description of a quantum gate must include not only the gate Hamiltonian but also the specification of both the input and output computational modes, $\hat a_\text{in/out} = \int\mathrm{d}\tau\,\Psi^{*}_\text{in/out}(\tau)\hat{a}_\tau$. In the presence of dispersion, we generically require $\hat a_\text{out} \neq \hat a_\text{in}$ even for a linear gate operation, and failure to choose an appropriate mode for the output can lead to loss of photons and hence fidelity. Of course, for a linear gate operation, one can always compute the correct output mode given the input mode using a linear scattering formalism, but this approach does not work for a nonlinear gate governed by a nonlinear multimode Hamiltonian. In this latter case, the problem devolves into numerical optimization of the input and output waveforms $\Psi_{\text{in/out}}(\tau)$, and even then, it is not guaranteed that there exists any input/output mode pair which allows for unit gate fidelity.

For the Kerr-phase gate, the intended action of the gate can be explicitly written in terms of photons in the input and output modes as
\begin{align}
    \label{eq:mapping-gaussian}
    \hat{U}_{\pi}\bigl[c_0\ket{0_\text{in}}+c_1\ket{1_\text{in}}+c_2\ket{2_\text{in}}\bigr]=c_0\ket{0_\text{out}}+c_1\ket{1_\text{out}}-c_2\ket{2_\text{out}},
\end{align}
where $\ket{n_\text{in/out}}=\frac{1}{\sqrt{n!}}\hat{a}_\text{in/out}^{\dagger n}(t)\ket{0}$ is the $n$-photon Fock state of the input/output mode. We also denote signal waveforms at intermediate times as $\Psi_a(\tau,t)$, so, for example, we have $\Psi_\text{in}(\tau)=\Psi_a(\tau,0)$ and $\Psi_\text{out}(\tau)=\Psi_a(\tau,t_\pi)$. Here, our aim is, by choosing appropriate input/output waveforms, to implement $\hat{U}_\pi$ as faithfully as possible given the fixed action of the waveguide $e^{-\mathrm{i}\hat{H}t_\pi}$. For this purpose, a reasonable approach would be to ensure at least perfect gate operation on the single-photon input, i.e., to fix the output mode according to
\begin{align}
\label{eq:single-photon-condition}
    \ket{1_\text{out}}=e^{-\mathrm{i}\hat{H}t_\pi}\ket{1_\text{in}},
\end{align}
which can be realized when $\Psi_a(\tau,t)$ is taken as the solution to
\begin{align}
\label{eq:signal-linear-dynamics}
    \mathrm{i}\partial_t\Psi_a=\hat{G}_a\Psi_a.
\end{align}
In this case, since the vacuum part evolves trivially, the only possible source of gate error is the action of $e^{-\mathrm{i}\hat{H}t_\pi}$ on the two-photon part $\ket{2_\text{in}}$, which can be characterized, e.g., by an error measure $\mathcal{D}=\Vert \ket{\psi_\text{out}}+\ket{2_\text{out}}\Vert=\sqrt{2(1+\mathrm{Re}\braket{\psi_\text{out}}{2_\text{out}})}$ with $\ket{\psi_\text{out}}=e^{-\mathrm{i}\hat{H}t_\pi}\ket{2_\text{in}}$.

A natural way to fulfill \eqref{eq:signal-linear-dynamics} is to set the input (and output) waveform to be an eigenmode of $\hat{G}^{(a)}$. In the absence of a temporal trap, however, these eigenmodes correspond to monochromatic cavity modes with weak nonlinearity due to their large mode volumes. Thus, we are motivated to consider nonstationary, pulsed solutions to \eqref{eq:signal-linear-dynamics} in order to increase the nonlinear coupling. To be concrete, we consider Gaussian waveforms which solve \eqref{eq:signal-linear-dynamics} as $\Psi_a(\tau,t)=e^{-\mathrm{i}\omega_{a,0}t}\Psi_\text{g}(\tau,t-t_\text{g})$ with
\begin{align}
    \Psi_\text{g}(\tau,t-t_\text{g}) & = \pi^{-1/4}\sqrt{\frac{\tau_\text{g}/\tau_\text{c}}{\tau^2_\text{g}/\tau_\text{c}^2+\mathrm{i}(t-t_\text{g})/t_\text{c}}} \nonumber \\
    & \ \ \ \ \  \times \exp\left(-\frac{\tau^2/\tau^2_\text{c}}{2(\tau_\text{g}^2/\tau_\text{c}^2+\mathrm{i}(t-t_\text{g})/t_\text{c})}\right). \label{eq:gaussian-waveform}
\end{align}
Here, $\tau_\text{g}$ is the pulse width, and $t_\text{g}$ characterizes the initial chirp, which we set to $t_\text{g}=t_\pi/2$ to minimize the maximum chirp during the gate operation. Where needed in the main text, we can similarly take the corresponding SH mode to be $\Psi_b(\tau,t)=e^{-\mathrm{i}\omega_{b,0}t}\Psi_\text{g}(\tau,\rho(t-t_\text{g}))$ with $\rho=\beta_{b,2}/\beta_{a,2}$.

\begin{figure}[t!]
\centering
    \includegraphics[width=1.0\columnwidth]{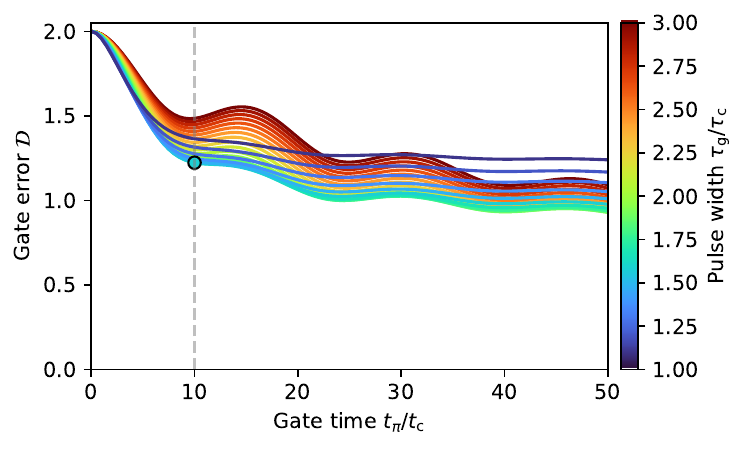}
    \caption{Gate performance of a Kerr-phase gate $\hat{U}_\pi$ implemented using Gaussian pulses shown for various gate times $t_\pi$ and pulse widths $\tau_\text{g}$. The gate error is measured as the distance $\mathcal{D}$ between the output state $\ket{\psi_\text{out}}=e^{-\mathrm{i}\hat{H}t_\pi}\ket{2_\text{in}}$ and the target state $-\ket{2_\text{out}}$. The solid circle represents the gate implementation with $\tau_\text{g}/\tau_\text{c}=1.5$ that maximizes the performance for a gate time of $t_\pi/t_\text{c}=10$. We assume $\omega_{b,0}-2\omega_{a,0}=0$ and $\beta_{a,2}=\beta_{b,2}/2$ for the simulation, and the system size $T$ is taken large enough compared to the pulse width $\tau_\text{g}$ such that there is no boundary effect.}
    \label{fig:gate-error}
\end{figure}

As described above, the gate error of this pulsed Kerr-phase gate is limited by its error acting on $\ket{2_\text{in}}$. Therefore, we present in Fig.~\ref{fig:gate-error} the error $\mathcal{D}$ for the input state $\ket{2_\text{in}}$ as a function of the gate time $t_\pi$ and pulse width $\tau_\text{g}$. While we do observe a slow improvement of the gate performance at longer gate times, perfect gate operation appears far from achievable. These gate errors are induced by undesired multimode nonlinear interactions with orthogonal modes, underscoring the challenges inherent to a traveling-pulse implementation of a nonlinear quantum gate. It is also worth mentioning that there exists a trade-off between the maximum temporal confinement and the rate of pulse dispersion in \eqref{eq:gaussian-waveform} (i.e., pulses with smaller width disperse faster), making it difficult to simultaneously achieve large temporal confinement and long interaction time.

On the other hand, in the presence of a temporal trap, we can take $\Psi_a$ to be a localized (bound) eigenmode of $\hat{G}_a$, which naturally leverages temporal confinement to enhance the nonlinear coupling. To see that this choice of $\Psi_a$ can realize effective single-mode dynamics and high-fidelity gate operations, let us consider the set of eigenmodes given by (for $u\in\{a,b\}$)
\begin{align}
    \hat{u}_m(t)=\int\mathrm{d}\tau\,e^{\mathrm{i}\lambda_{u,m} t}\Psi_{u,m}^*(\tau)\hat{u}_\tau,
\end{align}
where $\lambda_{u,m}$ and $\Psi_{u,m}(\tau)$ are the eigenvalue and the eigenmode of Eq.~(5) in the main text, respectively. The Hamiltonian Eq.~(4) of the main text rewritten in terms of these eigenmodes is
\begin{align}
    \hat{H}=\sum_{\ell mn}\frac{g_{\ell mn}}{2}e^{\mathrm{i}\delta_{\ell mn}t}\hat{a}_m^\dagger(t)\hat{a}_n^\dagger(t)\hat{b}_\ell(t)+\mathrm{H.c.}
\end{align}
with a nonlinear coupling tensor
\begin{align}
    g_{\ell mn}=r\int\mathrm{d}\tau\,\Psi_{b,\ell}^*(\tau)\Psi_{a,m}(\tau)\Psi_{a,n}(\tau)
\end{align}
and a phase-mismatch tensor $\delta_{\ell m n}=\lambda_{b,n}-\lambda_{a,\ell}-\lambda_{a,m}.$.  In the presence of a deep enough trap, we can realize $|\delta_{\ell m n}/g_{\ell m n}|\gg 1$, which strongly suppresses the nonlinear coupling unless special care is taken to make the process resonant. 

Specifically, with a temporal trap of the form $V_a(\tau) = V_b(\tau)/2= -(|\beta_{a,2}|/\beta_1\tau_0^2) \text{sech}^2(\tau/\tau_0)$ and $\rho=2$ considered in the main text, we have $\Psi_{a,0} = \Psi_{b,0}= (2\tau_0)^{-1/2} \sech(\tau/\tau_0)$ with a characteristic energy gap of $\Delta=\Delta_a=|\beta_{a,2}|/2\beta_1\tau_0^2$. By setting $\omega_{b,0}-2\omega_{a,0}=0$, we can set $\delta_{000}=0$ which brings the nonlinear interaction between fundamental eigenmodes (i.e., between the computational modes) to resonance. At the same time, leakage of photons from the computational modes are mediated by couplings of the form $g_{\ell00}$ and $g_{0mn}$, and it can be shown that $|\delta_{\ell00}|$ and $|\delta_{0mn}|$ are lower bounded by $\Delta$, thus suppressing leakage when $\Delta$ is large. As a result, photons are confined in the fundamental supermodes, where they experience single-mode dynamics described by an effective Hamiltonian
\begin{align}
\label{eq:single-mode-pulsed}
    \hat{H}\approx\frac{g}{2}(\hat{a}^2\hat{b}^\dagger+\hat{a}^{\dagger2}\hat{b}),
\end{align}
where we identify $g=g_{000}=\pi r/4\sqrt{2\tau_0}$, $\hat{a}=\hat{a}_0$, and $\hat{b}=\hat{b}_0$. As shown in the main text, the quantum dynamics under \eqref{eq:single-mode-pulsed} can be used to implement a high-fidelity Kerr-phase gate.

The emergence of the single-mode dynamics in the presence of a temporal trap is a generic phenomena and does not depend on the particular shape of the potential. For instance, under a more generic potential $V_a(\tau) = V_b(\tau)/2= -\alpha(|\beta_{a,2}|/\beta_1\tau_0^2) \text{sech}^2(\tau/\tau_0)$ and a dispersion $\rho=\beta_{2,b}/\beta_{2,b}$, we have~\cite{S_Manassah1990}
\begin{align}
    &\Psi_{a,0}=c_a\sech^{q_a}(\tau/\tau_0)&\Psi_{b,0}=c_a\sech^{q_b}(\tau/\tau_0),
\end{align}
where $c_u$ are normalization constants, and $q_a$ and $q_b$ are given as positive solutions of equations $q_a(q_a+1)/2=\alpha$ and $q_b(q_b+1)/2=2\alpha/\rho$, respectively (Notice that $\alpha=1$ and $\rho=2$ corresponds to the case discussed in the main text). When the phase-mismatch is set to $\omega_{b,0}-2\omega_{a,0}=(|\beta_{b,2}|q_b^2-2|\beta_{a,2}|q_a^2)/2\tau_0^2\beta_1$, the system Hamiltonian effectively reduces to the form \eqref{eq:single-mode-pulsed} with
\begin{align}
    g=\frac{r}{\pi^{1/4}\sqrt{\tau_0}}\frac{\Gamma(q_a+1/2)\Gamma(q_a+q_b/2)\Gamma^{1/2}(q_b+1/2)}{\Gamma^2(q_a)\Gamma(q_a+q_b/2+1/2)\Gamma^{1/2}(q_b)},
\end{align}
where $\Gamma(x)$ is the Gamma function.

\label{sec:shg}

\section{\uppercase{Nonlinear coupling literature comparison}}
\label{sec:a5}
In this section, we derive expressions relating the nonlinear coupling $g$ in the quantum model (Eq.~(2) in the main text) to experimentally measurable parameters, e.g., the normalized SHG conversion efficiency and the threshold power for optical parametric oscillation. These formulas are used to estimate the figure of merit $g/\kappa$ from devices presented in the literature. For the following, we consider a phase-matched $\chi^{(2)}$ resonator with a Hamiltonian $\hat{H} = \tfrac12 g (\hat{a}^{\dagger 2}\hat{b}+\hat{a}^2\hat{b}^\dagger)$ (analogous to Eq.~(\ref{eq:single-mode-pulsed})),
where we explicitly denote the reduced Planck constant by $\hbar$.  For both harmonics ($u\in\{a,b\}$), we denote the intrinsic and outcoupling decay rates by $\kappa_{u,\text{int}}$ and $\kappa_{u,\text{oc}}$, respectively. 

We first consider resonant SHG pumped by an external FH drive with power $P_\text{in}^{(a)}$, which can be modeled by a Hamiltonian term $\hbar\epsilon_a(\hat{a}+\hat{a}^\dagger)$ with $\epsilon_a=\sqrt{2\kappa_{a,\text{oc}}P_\text{in}/\hbar\omega_{a,0}}$. Under c-number substitution $\hat{a}\mapsto \alpha$ and $\hat{b}\mapsto\beta$, the classical dynamics of the fields follow
\begin{align}
    \mathrm{i}\partial_t \alpha&=g\alpha^{*}\beta-\mathrm{i}\kappa_a\alpha+\epsilon_a, & \mathrm{i}\partial_t \beta=\frac{g}{2}\alpha^2-\mathrm{i}\kappa_b\beta,
\end{align}
where $\kappa_u=\kappa_{u,\text{int}}+\kappa_{u,\text{oc}}$ is the total loss rate. In the undepleted-pump regime, the steady-state populations are
\begin{align}
    &|\alpha|^2=\frac{\epsilon_a^2}{\kappa_a^2},&|\beta|^2=\frac{g^2}{4}\frac{|\alpha|^4}{\kappa_b^2}.
\end{align}
Using the steady-state values, we can relate the normalized SHG conversion efficiency $\eta_\text{norm}=P_\text{out}/{P_\text{in}}^2$ with the output SH power $P_\text{out}=2\hbar\omega_{b,0}\kappa_{b,\text{oc}}|\beta|^2$ to the coupling coefficient $g$ as
\begin{align}
\label{eq:etanorm}
    \eta_\text{norm}=\frac{4g^2}{\hbar\omega_{a,0}}\frac{\kappa_{b,\text{oc}}}{\kappa_b^2}\left(\frac{\kappa_{a,\text{oc}}}{\kappa_a^2}\right)^2.
\end{align}
More general expressions for $\eta_\text{norm}$ in the case of the finite phase mismatch can be found in Ref.~\cite{S_Lu2020}.

Next, let us consider the scenario where the cavity is pumped by an external SH drive with power $P_\text{in}$. When the pump power is larger than some threshold value $P_\text{th}$, the system undergoes optical parametric oscillation (OPO). The external SH drive can be modeled by a Hamiltonian term $\hbar\epsilon_b(\hat{b}+\hat{b}^\dagger)$ with $\epsilon_b=\sqrt{2\kappa_{b,\text{oc}}P_\text{in}/\hbar\omega_{b,0}}$, leading to classical equation of motions
\begin{align}
    &\mathrm{i}\partial_t \alpha=g\alpha^{*}\beta-\mathrm{i}\kappa_a\alpha &\mathrm{i}\partial_t \beta=\frac{g}{2}\alpha^2-\mathrm{i}\kappa_b\beta+\epsilon_b.
\end{align}
The steady-state population of the FH mode takes a finite value under the condition
\begin{align}
    P_\text{in}\geq \frac{\hbar\omega_{b,0}\kappa_a^2\kappa_b^2}{g^2\kappa_{b,\text{oc}}}=P_\text{th},
\end{align}
which defines the OPO threshold power $P_\text{th}$. In particular, at critical coupling where $\kappa_{u,\text{int}}=\kappa_{u,\text{oc}}$, we have $P_\text{th}=2\hbar\omega_{b,0}\kappa_a^2\kappa_b/g^2$. Alternatively, some experimental results are reported in terms of the ``SHG saturation power'' related to the OPO threshold by $P_\text{sat}=4P_\text{th}$~\cite{S_Fuerst2010-opo}, which also allows us to calculate $g$ based on measurements of $P_\text{sat}$~\cite{S_Fuerst2010-shg, S_Ilchenko2004}.

\end{document}